\documentclass[pra,twocolumn,superscriptaddress]{revtex4}
\usepackage{graphicx}
\usepackage{amsmath, amssymb}
\usepackage{bm}
\usepackage{amsthm}
\usepackage{braket,mathrsfs}
\usepackage{CJK}

\def\w{\mathit{w}}
\def\be{\begin{equation}}
\def\ee{\end{equation}}
\def\ba{\begin{eqnarray}}
\def\ea{\end{eqnarray}}
\def\spp{\mathop{\mathrm{span}}}
\def\tr{\,\mathrm{tr}\,}
\def\la{\langle}
\def\ra{\rangle}
\def\d{\mathrm{d}}
\def\dE{\Delta_{\textrm E} E}
\def\DeltaE{\Delta_{\textrm Q} E}
\def\deltaE{\Delta_{\textrm {mc}} E}
\def\RE{\Delta_{\textrm C} E}

\begin{document}
    \title{Entropy for Quantum Pure States and Its Dynamical Relaxation}

     \author{Xizhi Han}
       \affiliation{International Center for Quantum Materials, School of Physics, Peking University, Beijing 100871, China}
    \author{Biao Wu}
      \email{wubiao@pku.edu.cn}
      \affiliation{International Center for Quantum Materials, School of Physics, Peking University, Beijing 100871, China}
      \affiliation{Collaborative Innovation Center of Quantum Matter, Beijing 100871, China}
      \affiliation{Wilczek Quantum Center, College of Science, Zhejiang University of Technology,  Hangzhou 310014, China}

    \begin{abstract}
        We construct a complete set of  Wannier functions which are localized at both
        given positions and momenta. This allows us to introduce the quantum phase space,
        onto which a quantum pure state  can be mapped unitarily. Using its probability distribution
         in quantum phase space, we define an entropy for a quantum pure state.
         We prove an inequality regarding the long time behavior of our entropy's fluctuation.
         For a typical initial state, this inequality indicates that  our entropy can relax dynamically
         to a maximized value and stay there most of time with small fluctuations. This result echoes
         the quantum H-theorem proved by  von Neumann in
        [Zeitschrift f\"ur Physik {\bf 57}, 30 (1929)].
        Our entropy is different from the standard von Neumann entropy, which is always zero
        for quantum  pure states. According to our definition,  a system always has bigger entropy
        than its subsystem even when the system is described by a pure state.
         As the construction of the Wannier basis can be implemented numerically,
         the dynamical evolution of our entropy is illustrated with an example.
     \end{abstract}
 \date{\today}
    \maketitle
\section{Introduction}
    Statistical mechanics, studying thermal properties of a many-body system from microscopic perspective, have gained huge success in the past century. However, the basic principles of statistical mechanics have not been fully understood; the establishment of micro-cannonical ensemble
has to rely on hypotheses\cite{HuangBook}.
Since microscopic particles --- elements of a macroscopic system --- are
governed by the Schr\"odinger equation, one feels obliged to address the problem
with quantum mechanics. Von Neumann was among  the first physicists trying to use
quantum mechanics to understand the basic principles of statistical mechanics. In a 1929 paper\cite{Von1929}, von Neumann proposed a method to construct commutable
macroscopic momentum and position operators and, therefore, quantum phase space.
Within this framework, he introduced an entropy for quantum pure state
and proved two theorems, which he called quantum ergodic theorem and
quantum H-theorem, respectively.
These results are remarkable advances in the establishment of the micro-canonical ensemble,
the foundation of statistical mechanics, without hypothesis. However, von Neumann's
beautiful results have been largely forgotten likely due to
misunderstanding\cite{vonNeumann2010commentary}.

Probably due to the developments in ultra-cold atomic gas
experiments\cite{Kinoshita2006,Schmiedmayer2013njp,Yukalov2011lpl},
we  have recently seen tremendous efforts to study the foundation of statistical
mechanics. Many new and beautiful results are obtained \cite{GemmerBook,Srednicki1994PRE,Typicality2006prl,Popescu2006Nphys,Sun2007pra,Rigol2008Nature,goldstein2010normaltypicality,linden2009thermalequilibrium,Reimann2007prl,reimann2008experimental,Kastner2012njp,Cho2010PRL,Ueda2011pre,Ji2011prls,Yukalov2011PLA,Sugiura2012prl,Snoke2012,Wang2012pre,Srednicki2012prl,Eisert2012prl,Emerson2013prl,zhuang2014,Goldstein2014short}.
These efforts have also led to renewed interest in von Neumann's forgotten work;
the English version of his paper is now available~\cite{vonNeumann2010qhtheorem}.
Von Neumann's quantum ergodic theorem has been re-exmained recently\cite{goldstein2010thermalequilibrium}.
In particular, a different version of quantum ergodic theorem was proved by
Reimann~\cite{reimann2008experimental,Short2011njp}. Reimann's ergodic theorem
does not involve any coarse-graining and can be subjected to
numerical study\cite{Zhuang2013pre}. In contrast, much less progress has been
made on the quantum H-theorem and the associated key concepts, such as
macroscopic momentum and position operators, and
entropy for quantum pure states, which were introduced in 1929.

In this work we define a different  entropy for quantum pure states
and study its long-time dynamical fluctuation in attempt to improve on
von Neumann's quantum H-theorem\cite{Von1929}. Von Neumann proved his theorem
with the following steps: ({\it i}) construct commutable macroscopic
position and momentum operators; ({\it ii}) define an entropy for pure quantum states
with coarse-graining; ({\it iii}) investigate the long-time behavior of the entropy.

We follow von Neumann's steps with new theoretical tools and perspectives.
For step ({\it i}), we use Kohn's  method~\cite{wkohn1973wannierfunctions} to
construct a complete set of  Wannier functions that are
localized both in position and momentum space.  Such a construction can be implemented numerically with great efficiency.  With these Wannier functions, we are able to construct commutable macroscopic position and momentum operators and, therefore, a
quantum phase space, which is divided into cells of size of the Planck constant
and each of these Planck cells is assigned a Wannier function.
The success of step ({\it i}) allows us to map unitarily a pure quantum state
onto the phase space.

We  accomplish step ({\it ii}) by defining
an entropy for a quantum pure state based on its probability distribution on the phase space.
Here we do not use coarse-graining used by von Neumann in the context of macroscopic
observables.  For our entropy,  the total system always has a larger entropy than its
subsystems even if the total system is described by a quantum pure state. This is not the case for
the conventional von Neumann's entropy for mixed states.

For step ({\it iii}),  we introduce an ensemble entropy for a pure state and
prove an inequality regarding the dynamical fluctuation of our entropy, which is
similar to von Neumann's quantum H-theorem. This inequality includes a constant
$C$ that characterizes the correlation of probability fluctuations between different Planck
cells. When the correlation is small,  the inequality dictates that
our entropy relax dynamically to the ensemble entropy
and stay at this value most of time with small fluctuations for macroscopic systems.
Our analysis shows that $C$ is small as long as the energy shell of microcanonical ensemble
is not too narrow and not sporadically populated.  As a result,
a better understanding of the microscopic  origin of the second law of
thermodynamics is achieved.  The long-time dynamical  evolution of our entropy
is illustrated numerically with an example.

\section{Quantum phase space}

To establish quantum phase space, von Neumann proposed  to construct a macroscopic position operator $\bm{Q}$ and a macroscopic momentum operator $\bm{P}$ that satisfy\cite{vonNeumann2010qhtheorem}
    \begin{eqnarray}
        [\bm{Q}, \bm{P}] = 0 \label{eq:hyp1}\,,\\
        \bm{Q} \sim \bm{q}, \quad \bm{P} \sim \bm{p}\,, \label{eq:hyp2}
    \end{eqnarray}
where $\bm{q}$ and $\bm{p}$ are usual microscopic position and momentum operators, respectively, that have the commutator $[\bm{q}, \bm{p}] = i\hbar$.  Eq.\;(\ref{eq:hyp2}) indicates that the macroscopic  position and momentum operators are not identical but close to their microscopic counterparts.
Mathematically it is equivalent to finding a complete set of normalized orthogonal wave functions
$\{\w_j\}$ localized in both position and momentum spaces.
The macroscopic position and momentum operators can then be expressed as
\begin{eqnarray}
    \bm{P} = \sum_j |\w_j\rangle\langle \w_j|\bm{p}|\w_j\rangle\langle \w_j|\,, \\
    \bm{Q} = \sum_j |\w_j\rangle\langle \w_j|\bm{q}|\w_j\rangle\langle \w_j|\,.
\end{eqnarray}
 Eq.\;(\ref{eq:hyp2}) implies that the $i^\textrm{th}$ order central moments
 \ba \Delta^{(i)} p_j \equiv \langle \w_j|(\bm{p} - \la\bm{p}\ra_j)^i|\w_j\rangle^{1/i} \label{spreadp}\\
 \Delta^{(i)} q_j \equiv \langle \w_j|(\bm{q} - \la\bm{q}\ra_j)^i|\w_j\rangle^{1/i} \label{spreadq}\ea
 should be relatively small for  all $i\ge 2$. $\la f\ra_j$ denotes $\la w_j | f | w_j \ra$.  For convenience, we often denote $\Delta^{(2)}$ simply by $\Delta$.

For one-dimensional system in which $\bm{q} \equiv \bm{x}$,
$\bm{p} \equiv \hbar \bm{k} = - i \hbar \partial_x$, von Neumann
proposed to find $\{\w_j\}$ by Schmidt orthogonalizing a set of Gaussian
wave packets of width $\zeta$\cite{Von1929}
\begin{equation}
\label{gaussian}
g_{j_x,j_k}\equiv\exp\big[-\frac{(x - j_x x_0)^2}{4 \zeta^2} + i j_k k_0 x\big]\,,
\end{equation}
where $j_x,j_k$ are integers. When $x_0 \,k_0=2\pi$, this set of Gaussian packets are
complete. We are at liberty to choose $x_0$, $k_0$, and $\zeta$ as long as
$x_0 \,k_0=2\pi$ is satisfied.
Unless otherwise specified, parameters are chosen as  $x_0 = 1$, $k_0 = 2 \pi$ and $\zeta = (2 \pi)^{-1}$.

This method, which is called ``cumbersome" by von Neumann
himself~\cite{vonNeumann2010qhtheorem}, suffers from two major drawbacks.
First, it is not feasible numerically due to its high computational cost and
sensitivity to the order of the orthogonalization procedure.
Secondly, von Neumann argued \cite{Von1929} that the existence
of $\bm{P}$ and $\bm{Q}$ corresponds to the fact that
the position and momentum can be measured simultaneously in macroscopic measurements.
As  there is no difference among measuring positions at different spatial points,
we expect that the constructed $\{\w_j\}$ have spatial translational symmetry.
However, the wave packets constructed with von Neumann's method have no such symmetry.

\subsection{Wannier Basis}
Kohn suggested a method to construct Wannier functions out of Gaussian wave
packets~\cite{wkohn1973wannierfunctions}.  We adapt Kohn's approach to
orthogonalize the Gaussian packets in  Eq.(\ref{gaussian}) and construct a
complete set of Wannier functions $\{\w_j\}$ whose translational symmetry
is guaranteed. The detailed procedure of construction  is elaborated as follows.
    \begin{enumerate}
      \item Choose an initial set of localized wave packets such as
      the Gaussian wave packets $g_{j_k}(x)\equiv g_{0,j_k}(x)$ in Eq.\;(\ref{gaussian}).
      Find their Fourier transform $\tilde{g}_{j_k}(k) \equiv \mathcal{F}\{g_{j_k}(x)\} \equiv \frac{1}{\sqrt{2\pi}} \int g_{j_k}(x)
      \mathrm{e}^{-i k x} \;\mathrm{d} x$.

      \item At a fixed $k$, for every $j_k$,   $(\tilde{g}_{j_k}(k + 2 n \pi))_{n \in \mathbb{Z}}$ is a
      normalizable vector; we denote it by  $u_{k, j_k}(n)$.  Apply Schmidt orthogonalization procedure
       $v_0 = u_0$ (the subscript $k$ is omitted), normalize $v_0$, $v_1 = u_1 - (u_1, v_0) v_0$, normalize $v_1$, and repeat for $u_2$, $u_3,\cdots$. We eventually
       get an orthonormal basis $\{v_{k, j_k} \in l^2(\mathbb{Z})\}_{j_k \in \mathbb{Z}}$.
       Define $\tilde{\w}_{j_k}(k + 2 n \pi) \equiv v_{k, j_k}(n) / \sqrt{2 \pi}$.

      \item For every $k$ (discrete in numerical calculations) on $[0, 2 \pi)$,  repeat step 2. According to Proposition \ref{thm1} in Appendix A, $\w_{j_x,j_k}(x)\equiv\w_{j_k}(x - j_x)$
      ($\w_{j_k}$ is the Fourier transform of  $\tilde{\w}_{j_k}$) are orthonormal.
      $\{w_j\}$ is the desired orthonormal basis ($j=(j_x, j_k)$).
    \end{enumerate}

\begin{figure*}
            \includegraphics[width=0.35\textwidth]{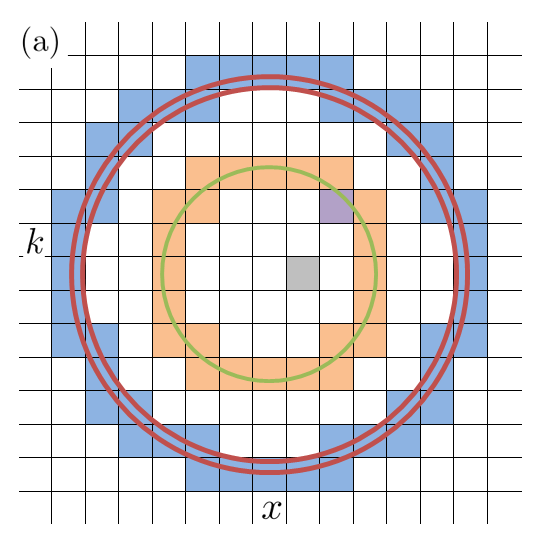}
            \includegraphics[width=0.45\textwidth]{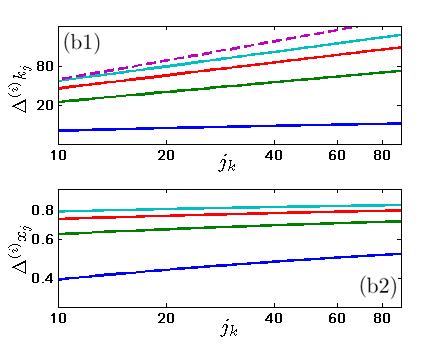}\\
            \includegraphics[width=0.41\textwidth]{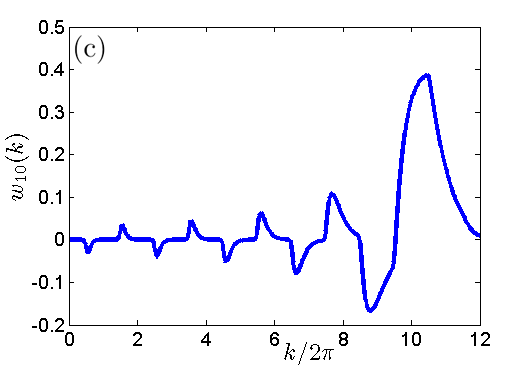}
            \includegraphics[width=0.43\textwidth]{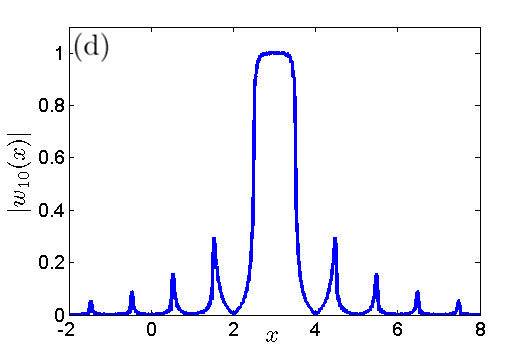}
        \caption{Illustration of quantum phase space. (a) Schematic plot of quantum phase space  (b)
        Spreads of Wannier functions as a function of $j_k$. Both axes are in logarithmic
        scales. Solid lines are for  $i = 8, 6, 4, 2$, from top to bottom, respectively.
        The dashed line of slope 1 is drawn to show that all solid lines have slope less than 1.
        (c) Wannier function $\w_{3,10}$ is shown localized near $k = 10\cdot 2 \pi$. (d) Wannier function $\w_{3,10}$ is shown localized near $x = 3$.}
        \label{phspace}
    \end{figure*}

We have thus established a quantum phase space which is different from
the classical phase space: (1) It is divided into phase cells  of size
Planck constant $h$ (for one dimensional system) as illustrated in
Fig. \ref{phspace} (a); we call such a cell Planck cell for brevity.
(2) Each Planck cell is assigned a Wannier function $w_j$, which is
localized near site ($x=j_x$, $k=2j_k\pi$).
We are now able to map a pure wave function unitarily onto phase space. There has been tremendous efforts to
formulate quantum mechanics in phase space based on Wigner's quasi-distribution function and
Weyl's correspondence~\cite{qphasespace}. However, Wigner's
quasi-distribution is not positive-definite and cannot be interpreted
as probability in phase space. According to our construction, for a wave function $\psi$,
$|\braket{\psi|\w_j}|^2$ is its probability
at Planck cell $j$  as $\{\w_j\}$ is a set of complete orthonormal basis.

The generalization to higher dimensions is straightforward. With
the one-dimensional $\{\w_j(x)\}$ that we have constructed,  we simply define
\begin{equation}
\w_{j_1 j_2 \ldots j_n}(x_1, x_2, \ldots, x_n) \equiv \w_{j_1}(x_1) \w_{j_2}(x_2) \ldots \w_{j_n}(x_n)\,. \label{highd}
\end{equation}
Then $\{\w_{j_1 j_2 \ldots j_n}\}$ is the localized orthonormal basis for an $n$-dimensional system.

Numerical results of one-dimensional Wannier functions  are provided in Fig.\;\ref{phspace}.
A Wannier function localized near $(x=3,k=20\pi)$ is
plotted in the $k$ and $x$ spaces, respectively, in Fig.\;\ref{phspace}(c) and (d).
This Wannier function is obtained with the above procedure
using the Gaussian wave packets $g_{j_x,j_k}$ as initial functions.  And
the order of Schmidt orthogonalization in our procedure is chosen to be
$j_k = 0, 1, -1, 2, -2, \ldots$. The result does not sensitively depend on the order.

Our numerical computation finds that the Wannier function spreads out slowly with
increasing momentum $k$. From Fig.\;\ref{phspace} (b) we can see that
both $\Delta^{(i)} k_j$ and $\Delta^{(i)} x_j$, which characterize the spreads of
the Wannier function, diverge as $j_k$ increases; $\Delta^{(i)} x_j$ appears to grow
more slowly.
Actually, it can be proved that the product of $\Delta x_j\cdot \Delta k_j$ diverges
as $j_k$ increases no matter what initial wave packets are chosen (see Appendix B).
This divergent behavior of $\Delta x \cdot \Delta k$ is called strong uncertainty
relation\cite{bourgain1988remark}.

However, the divergence is not very severe.  As shown in Fig.\;\ref{phspace} (b)
where both axes are in logarithmic scales, all the growth slopes are much less than one.
Therefore,
all orders of the relative spreads $\Delta^{(i)} x_j / 2 \pi j_k$ and $\Delta^{(i)} k_j/ 2 \pi j_k$
fall to zero quickly as $j_k$ increases.
This suggests that for the one-dimensional system, the requirement (\ref{spreadp}) and (\ref{spreadq}) are satisfied in the sense
\be \lim_{\la p \ra_j / p_0 \to \infty} \frac{\Delta^{(i)} p_j}{\la p \ra_j} = \lim_{\la p \ra_j / p_0 \to \infty} \frac{p_0 \cdot \Delta^{(i)} q_j }{\la p \ra_j q_0} = 0\,,
\label{localization}
\ee
where we have used $p=\hbar k$, $q=x$, and
$\la p\ra_j \approx j_k p_0$.

\subsection{Quantum Energy Shell}
In classical phase space, there is an important concept of energy
surface,  where the dynamics of an isolated system is confined.
Energy surface, which is of no width,  is no longer valid in
the quantum phase space which consists of cells of finite size.
However,  a similar concept,  energy shell of finite width, can be introduced.
For this purpose, we need to first show that each of our Planck cells is localized in energy
for most of the macroscopic systems of physical interest.

For an isolated system of fixed number of particles $N \gg 1$ with Hamiltonian $H(\bm{p}, \bm{r})$ where $\bm{p}$ and $\bm{r}$ are $3N$-dimensional vectors, define $p_C$ as the typical magnitude of momentum of any particle and $r_C$ as the typical length scale on which $H$ changes relatively significantly. For example, $r_C$ can be  the mean free path of a particle or
the characteristic scale of the external potential. We define the index
\be I \equiv p_C r_C / h \ee
In this work we focus on the cases where $I$ is considerably large.

We expect that the quantum phase space is reduced to the classical phase space in
the limit $I \to \infty$ in the sense that the relative size of a Planck cell  and
the relative spreads of the Wannier functions tend to zero.  This is indeed the case.
We construct Planck cells defined by $p_0 = p_C / \sqrt{I}$ and $r_0 = r_C / \sqrt{I}$.
We immediately have $p_0 / p_C = r_0 / r_C = I^{-1/2} \to 0$ in the limit $I \to \infty$.
Suppose that $j_C$ is the momentum index such that $\la p\ra_{j_{C}} \approx p_C$.
For a typical Planck cell $j$ whose $|j_k| \lesssim j_C$, we have
according to Eq.\;(\ref{localization})
    \be
    \frac{\Delta^{(i)} p_j} {p_C} \lesssim \frac{\Delta^{(i)} p_{j_C}}{p_C} \to 0 \,,
    \ee
    and similarly,
    \be
    \frac{\Delta^{(i)} r_j }{r_C} = \frac{p_0}{r_0} \frac{\Delta^{(i)} r_j }{p_C} \lesssim \frac{p_0}{r_0} \frac{\Delta^{(i)} r_{j_C} }{p_C} \to 0\,,
    \ee for $i = 2, 3, \dots$ in the limit $I \to \infty$.
We obtain  the desirable picture, the quantum phase space becoming the classical phase space
as $I \to \infty$. We thus call  $I \to \infty$ classical limit.
We will continue to use this choice of $p_0$ and $r_0$ in the following discussion.

Now we are ready to show that indeed our Wannier functions are localized in energy. To avoid cumbersome partial derivatives and summations, we illustrate the point with single-particle one-dimensional potential $V(x)$; the case of kinetic energy and multi-particle systems should be essentially the same. For a typical Planck cell $j$,
we expand $V$ at $\la x\ra_j$ where $|w_j\rangle$
is localized and compute its relative spread
\be
\frac{\Delta V^2}{V_0^2}=\left\la\frac{(V - V_0)^2}{V_0^2}\right\ra_j = \sum_{i = 2}^\infty a_{j,i}
\left(\frac{\Delta^{(i)} x_j}{r_C}\right)^i
\ee
where $V_0 \equiv V(\la x\ra_j)$. As $V$ varies on the scale $r_C$ ,
it is easy to see that  $a_{j, i} = O(1)$. Therefore, the relative spread $\Delta V^2/V_0^2$
tends to zero in the classical limit $I \to \infty$.

As our Wannier functions are localized in energy, when we map an energy
eigenstate  $\ket{\phi_\alpha}$ with eigen-energy $E_\alpha$
to the quantum phase space, only the Planck cells
with their energies $E_j=\braket{w_j|H|w_j}\sim E_\alpha$ are significantly occupied.
We say that energy eigenstate $\ket{\phi_\alpha}$ crosses Planck cell $j$ when $\la w_j | \phi_\alpha \ra$ is significantly non-zero.
As a result, we can define
an energy shell $a$ of energy interval $[E_a, E_a+\Delta E_a]$ as  a set of phase
cells $\w_{j}$'s such that $\langle\phi_\alpha|\bm{\Delta}_a|\phi_\alpha\rangle\sim 1$
when $E_\alpha\in [E_a, E_a+\Delta E_a]$.    The projection operator
$\bm{\Delta}_a \equiv \sum_{j=1}^{N_a} |\w_{j}\rangle\langle\w_{j}|$,
where $N_a = \tr \bm{\Delta}_a$ is the number of Planck cells in energy shell $a$.
Energy shell $a$ is said to be significantly occupied by a quantum state $\psi(t)$ when $\overline{\la\psi(t)|\bm{\Delta}_a|\psi(t)\ra}$ is considerably larger than zero.


We draw the quantum phase space schematically in Fig.\;\ref{phspace}(a),
where  squares are for Planck cells and circles represent eigen-energies.
Two energy shells are illustrated: one with blue Planck cells and the other with orange Planck cells.
Each energy eigenstate  may cross many Planck cells; at the same time, one Planck cell
can be crossed by many energy eigenstates. The purple Planck cell is in the orange energy shell while the gray one is in neither shell colored.


 \section{Hierarchy of Energy Scales}
In this section we examine the energy scales involved and establish a hierarchy
among them. It will become clear later that these energy scales and
their hierarchy play crucial roles in regulating the long time dynamics of the system.

One energy scale is $\dE$, the typical difference between adjacent eigen-energies.
The typical energy uncertainty $\DeltaE$ in a Planck cell is another energy scale.
For a typical Planck cell $j$, we have
\be
\DeltaE = \big[\sum_{\alpha} (E_\alpha - E_j)^2  |\la w_j | \phi_\alpha \ra|^2\big]^{\frac{1}{2}}\,.\label{DeltaE}
\ee
For a quantum system  with large number of particles $N \gg 1$,  it should be expected
that though $w_j$'s are localized in energy, eigenstates that cross every Planck cell
are numerous. To see this, we note that the density of state $\rho(E)$ grows
exponentially while $\DeltaE$  increases polynomially as $N \to \infty$.
Therefore, for a typical many-particle system, we have $\dE \ll \DeltaE$.

Consider a general quantum state
$
\ket{\psi(t)}=\sum_\alpha c_\alpha (t) \ket{\phi_\alpha}\,,
$
and denote $\wp_\alpha\equiv|c_\alpha (t)|^2$.  For this quantum state,
there exists an energy scale $\RE$  defined as
\be
\RE = \big[\sum_{j, \alpha} (E_\alpha - E_j)^2 \wp_\alpha |\la w_j | \phi_\alpha \ra|^2\big]^{\frac{1}{2}}\,,\label{RE}
\ee
where $E_j\equiv\braket{w_j|H|w_j}$ is the average energy of Planck cell $j$.  We call
$\RE$ the correlation energy scale. As we will show later, only the Planck cells
which are separated by energy less than $\RE$  are correlated. A comparison between
Eq.\;(\ref{DeltaE}) and Eq.\;(\ref{RE}) indicates that we have $\DeltaE\sim\RE$ for
a typical quantum state.

Many properties, in particular macroscopic properties of a system,  are not sensitive
to the details of a quantum state. Since $\dE \ll \DeltaE$, we define a
smoothed function over energy scale $\DeltaE$ as follows
\be
 \la f_\alpha \ra_s(E) \equiv \sum_{|E_\alpha - E| < \DeltaE} f_\alpha \big/ \sum_{|E_\alpha - E| < \DeltaE} 1 \,,
 \ee
For example,  $\la \wp_\alpha\ra_s(E)$ is the smoothed probabilities of the quantum state
$\psi$ at $E$. We can now introduce another energy scale  $\deltaE$ on which
$\la \wp_\alpha\ra_s$  can be regarded as constant. This energy scale $\deltaE$
indicates the width of the energy shell which is significantly occupied by $\psi$.
In this work we focus on the quantum state such that the following hierarchy
of magnitudes is satisfied,
\be
\dE \ll \DeltaE \sim \RE \ll \deltaE \ll E \label{hierarchy}\,,
\ee
where $E\equiv\braket{\psi|H|\psi}$.  For a quantum state prepared in real experiments for
a many-body system, both $\deltaE $ and $E$ are of macroscopic size while
$\DeltaE$ and $\RE$ are microscopic. Therefore, the hierarchy in Eq.\;(\ref{hierarchy})
are readily satisfied in real experiments.

In textbooks on quantum statistical mechanics\cite{HuangBook}, the  micro-canonical ensemble
is established on an energy shell of width  $\deltaE \ll E$. Usually no lower bound is
given for $\deltaE$. Here we see that it should have a quantum lower bound of $\DeltaE$,
which will be shown later to play a key role to guarantee the equilibration of the system.

Finally, we assume that the eigenstates are not highly concentrated in the highly occupied
energy shell $[E, E+\deltaE]$. Mathematically, this means that
the density of states $\rho(E)$ satisfies
 \be
  \int_{E_j - \DeltaE}^{E_j + \DeltaE} \d E \;\rho(E) \ll
  \int_{E_j - \deltaE}^{E_j + \deltaE} \d E \;\rho(E)\,. \label{inevitable}
 \ee
 Despite a few exceptions(flat band etc.),  this assumption is not strong and should be satisfied by most of the macroscopic systems in high energy states.

 \section{Entropy for Pure Quantum State and an inequality for its Fluctuations}
    As we can now map a wave function unitarily to the quantum phase space, we can use
    its probability distribution in the phase space to define an entropy.
    For a pure quantum state $\psi(\bm{r})$, we define its entropy as
    \begin{equation}\label{eq:basisentropy}
    S_{\w}(\psi) \equiv -\sum_{j} \langle\psi|\bm{W}_j|\psi\rangle \ln \langle\psi|\bm{W}_j|\psi\rangle
    \end{equation}
    where $\bm{W}_j \equiv |\w_j\rangle\langle \w_j|$ is the projection to Planck cell $j$ characterized by
    Wannier function $\w_j(\bm{r})$.

Consider an isolated quantum system described by $\psi$.  As this state
evolves with time according to the Schr\"odinger equation, its entropy $S_{\w}(\psi)$ will evolve in time.
Will the entropy increase and eventually approach a maximum in accordance with the second law of thermodynamics? The answer is yes for a large class of quantum systems in the sense established by
von Neumann in 1929\cite{Von1929}. In the 1929 paper, von Neumann introduced an entropy for pure quantum
states; he then proved an inequality concerning the long time dynamical behavior of this entropy.
According to this inequality,  if the system starts with a low entropy state, the system will evolve
into high entropy states and stay there almost all the time with small fluctuations. Von Neumann
called this inequality quantum H-theorem.  We will prove a similar inequality in this section.

As the system evolves,
the probability in each Planck cell $j$ will change with time ($\hbar = 1$)
\begin{eqnarray}
&&\wp_j(t)\equiv\braket{\psi(t)|\bm{W}_j|\psi(t)}\nonumber\\
&&=\sum_{\alpha,\beta}\braket{\psi(0)|\phi_\alpha}\braket{\phi_\alpha|\bm{W}_j|\phi_\beta}\braket{\phi_\beta|\psi(0)} e^{i(E_\alpha-E_\beta)t}\,.\nonumber\\
\end{eqnarray}
We define $\overline{\wp}_j$ as the long time averaging of $\wp_j(t)$ and introduce a corresponding entropy
\be
S_E(\psi)\equiv-\sum_j \overline{\wp}_j\ln \overline{\wp}_j\,. \label{def:se}
\ee
We 
call it ensemble entropy for pure state $\psi$.
The ensemble entropy $S_E$ does not change with time. We find that under some reasonable conditions, the entropy $S_w(\psi)$ will approach
$S_E(\psi)$ and stay close to it almost all the time with small fluctuations. First we present a rather universal inequality concerning the long time behavior of our entropy, which will imply the equilibration of our entropy under reasonable conditions.
We leave details of the proof to Appendix C; the inequality is as follows.
\newtheorem*{hthm}{Theorem }
    \begin{hthm}
    For a quantum system  governed by a Hamiltonian whose eigenvalues satisfy the following conditions 1, 2 and 3, and for every $j$, $0 \leq \overline{\wp}_j \leq 1 / \mathrm{e}$, we have
    \be
    \frac{\overline{(S_w(\psi(t)) - S_E)^2}}
    {S_E^2} \leq C    + \frac{8} {S_E} + \frac{4}{S_E^2}\,,
    \label{htheorem}
    \ee
    where \be C \equiv \sum_{j, j'} C_{j j'}\left(\overline{\wp}_j \ln \overline{\wp}_j\right)\left(\overline{\wp}_{j'} \ln \overline{\wp}_{j'}\right) \big/ \big(\sum_j \overline{\wp}_j \ln \overline{\wp}_j\big)^2 \label{average}\ee and $C_{j j'} \equiv \overline{(\wp_j(t) - \overline{\wp}_j) (\wp_{j'}(t) - \overline{\wp}_{j'})} \big/ \overline{\wp}_j \overline{\wp}_{j'}$.
\end{hthm}
The three conditions are
     \begin{itemize}
      \item {\it Condition 1:}\; $E_\alpha = E_\beta \Rightarrow \alpha = \beta$;
      \item {\it Condition 2:}\; $E_{\alpha} - E_{\beta} = E_{\alpha'} - E_{\beta'},
      \alpha \neq \beta \Rightarrow \alpha = \alpha', \beta = \beta'$;
      \item {\it Condition 3:}\; $E_\alpha + E_\chi - E_\beta - E_\gamma = E_{\alpha'} + E_{\chi'} - E_{\beta'} - E_{\gamma'}$, $\{\alpha, \chi\} \cap \{\beta, \gamma\} = \emptyset$
      $\Rightarrow \{\alpha, \chi\} = \{\alpha', \chi'\}$ and $\{\beta, \gamma\} = \{\beta', \gamma'\}$.
    \end{itemize}
Condition 1 and 2 are commonly used~\cite{Von1929,reimann2008experimental,Short2012njp}, representing no degeneracies of energies and energy gaps, respectively.
Condition 3 implies differences between energy gaps are also distinct. From the random matrix theory\cite{StockmannBook}, we believe condition 3 should be satisfied by most non-integrable systems; as a result, the inequality should hold for majority of quantum systems. These three conditions have a close connection with moments of $\wp_j(t)$ statistically, i.e. $\overline{\wp_j(t)}$, $\overline{\wp_j(t)^2}$ and $\overline{\wp_j(t)^4}$. For example, with condition 1, we have
\be
\overline{\wp_j}=\sum_\alpha c_{\alpha j}^*c_{\alpha j}\,.
\ee
where $c_{\alpha j}=\braket{\psi(0)|\phi_\alpha}\braket{\phi_\alpha|w_j}$. For the rest of
details, please see Appendix C.

We now discuss the physical interpretation of $C$, $C_{j j'}$ and the inequality. Clearly, $0 \leq |C_{j j'}| \leq 1$ signifies the fluctuation correlation between Planck cells $j$ and $j'$; $C$ can be regarded as some kind of averaging over $C_{j j'}$ with weight $-\overline{\wp} \ln \overline{\wp}$. Hence $C$ characterizes the averaged fluctuation correlation between cells.
With such understanding, the inequality can be understood intuitively: when $S_E$ is large, that is the probability distribution spreads over many Planck cells, the correlation of $\wp$ between the majority of Planck cells are small;  the total entropy $S_w $ undergoes small fluctuations
most of time.  In these situations,   the inequality (\ref{htheorem})  implies a
quantum H-theorem similar to von Neumann's.

Indeed we can  demonstrate that $S_E$ is large and $C$ is small under the following two conditions:
\begin{itemize}
\item The hierarchy (\ref{hierarchy}) and the assumption (\ref{inevitable})  hold.
\item For significantly occupied energy shells, the occupancy rate
\be
R \equiv \la \wp_\alpha \ra^2_s  / \la \wp_\alpha^2 \ra_s
\ee
is high.
\end{itemize}
$R$ signifies the fluctuation of $\wp_\alpha$: if all eigenstates are equally occupied, $R = 1$;
if only one of $N_c$ consecutive eigenstates is occupied, $R = 1 / N_c$.

\paragraph{Estimate of $S_E$} We can show  (see Appendix D)
\be
S_E^{\max} - S_E \lesssim -\ln R \,, \label{SEmax}
\ee
where
\be
S_E^{\max} = -\int_{-\infty}^\infty \d E\; \rho(E) \la \wp_\alpha \ra_s(E) \ln \la \wp_\alpha \ra_s(E)\,.\ee
By Jensen's inequality $ S_E^{\max} \geq \ln d_{\textrm{eff}} $
where $d_{\textrm{eff}}^{-1} \equiv \sum_\alpha \wp_\alpha^2$ is the effective number of eigenstates occupied~\cite{Short2011njp}.  $d_{\textrm{eff}}$ can certainly also
be regarded as the microscopic states occupied in a macroscopic quantum state.
For a quantum state prepared in real experiments,
$d_{\textrm{eff}}$ is a very large number~\cite{reimann2008experimental,Short2011njp}.
When $R$ is reasonably high, $R\sim 1$,
we have $S_E\approx S_E^{\max}\geq \ln d_{\textrm{eff}}$.
Therefore, $S_E$ is indeed very large.

\paragraph{Estimate of C} As our Wannier functions are localized in energy as discussed in Section II, Planck cells $i$ and $j$ far apart are not likely to share energy eigenfunctions (that is, for energy eigenstate $\phi_\alpha$, $\la w_i | \phi_\alpha \ra$ and $\la w_j | \phi_\alpha \ra$ are not significant simultaneously); thus with condition 1 and 2, $\wp_i$ should not be considerably correlated with $\wp_j$. When hierarchy (\ref{hierarchy}) holds, $\DeltaE \ll \deltaE$, pairs of Planck cells not significantly correlated should be the majority in Eq.\;(\ref{average}). As a result, $C$ should be small. In fact we estimate (see Appendix D)
\be C \lesssim \DeltaE / \deltaE \label{estimateC} \,.\ee

When $S_E$ is maximized with $C\approx 0$,  the inequality (\ref{htheorem}) shows that
the relative fluctuation of $S_w$ away from $S_E$ is small when it is averaged over a long time.
This means that when the system starts with a low entropy state, it will relax dynamically to
states whose entropies are very close to $S_E$. Otherwise the inequality would be violated. Note that
it is possible that the system can evolve into a state whose entropy is far away from $S_E$.
When this happens, the system will relax dynamically back in a short time to states whose entropies
are high and close to $S_E$. This reminds us the Poincar\'e recurrence in classical dynamic systems.
So, the morale is the same for both quantum and classical dynamics:
due to the time reversal symmetry inherently possessed by both quantum and classical systems,
it is impossible to rule out that the system evolves dynamically to a lower entropy state. However,
with conditions above we can assert that the large deviation from
the maximized entropy is possible only rarely in quantum dynamics.

As the quantum system  equilibrates, not only its entropy reaches its maximum, other
observables such as momentum or density distribution also settle.
In our definition of entropy, it is clear when the entropy reaches its maximum, $\la w_j |\psi\ra$ can acquire distinct phase factors while not affecting the total entropy. When $\la w_i | \bm{p} | w_j \ra$ is small (relatively) for $i \neq j$, $\la \psi | \bm{p} | \psi \ra$ does not significantly depend on those phase factors, either. In macroscopic systems, if Planck cell $i$ and $j$ are close to each other (with $|p_i - p_j|$ much less than $p \equiv (p_i + p_j) / 2$), $\bm{p}$ can be regarded as a constant on the cells thus $\la w_i | \bm{p} | w_j \ra \approx p \la w_i | w_j \ra = 0$ for $i\neq j$; if Planck cell $i$ and $j$ are far apart, their overlapping is small and as a result, $\la w_i | \bm{p} | w_j \ra$ is relatively small. Similar argument applies to other observables (such as $\bm{x}$) as long as the observable varies on a scale much larger than $\Delta p$ and $\Delta x$ of Wannier functions. The inequality proved by Riemann for the fluctuations of observables
~\cite{reimann2008experimental,Short2011njp} is also
an indication that observables should equilibrate when the entropy approaches its maximum value.




 \begin{figure}[h!]
            \includegraphics[width=0.45\textwidth]{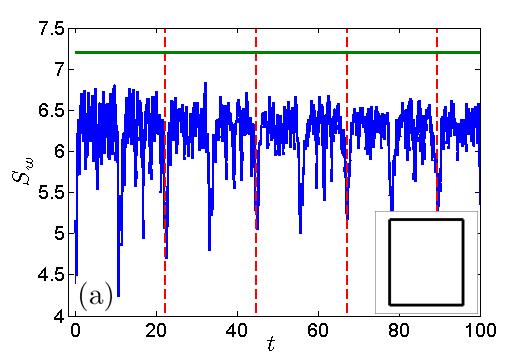}
            \includegraphics[width=0.45\textwidth]{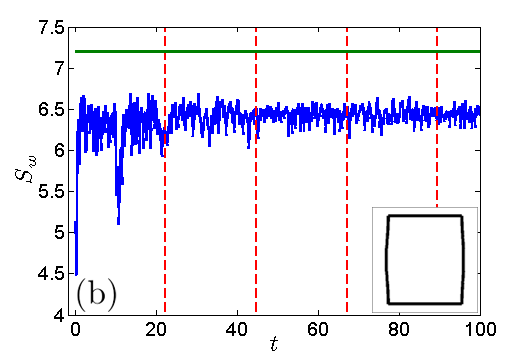}
            \includegraphics[width=0.45\textwidth]{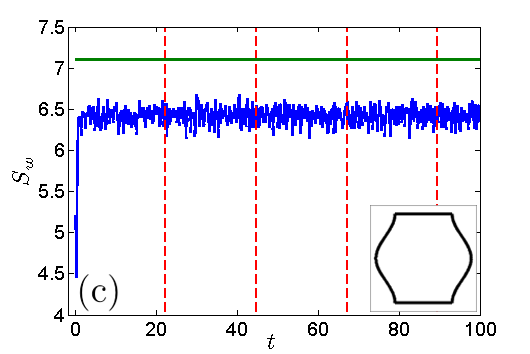}
        \caption{Time evolution of $S_w(\psi(t))$ for three different ripple billiards. The initial state is
        a moving Gaussian wave packet. The dashed red lines indicate the theoretical recurrence time for a square infinite potential well of size $2 b \times 2 b$.  The green lines are the ensemble entropy $S_E$.
        The three ripple billiards shown in the insets are characterized by
        $\epsilon =a/b= 0.25\%, 2.5\%, 25\%$, respectively. }
        \label{ripple}
    \end{figure}

As we are able to compute the Wannier functions ${\w_j}$ numerically, the entropy for quantum
pure states and the relaxation of our entropy towards a maximum can now be illustrated with a concrete example.
We are trying  to answer whether a macroscopic many-body quantum system
can equilibrate dynamically. However, as we have seen in this work and in
many others' work\cite{Von1929,reimann2008experimental},
the conclusion relies on only the structure of eigen-energies of the system (degeneracy,
energy gaps, etc.), which are shared by both single-particle and many-body systems according to
the random matrix theory\cite{StockmannBook}. This means that in many situations it  is sufficient to use
single-particle systems to illustrate entropy for pure states and the quantum H-theorem.

We choose to use ripple billiard with which we are very familiar.  The ripple billiard
is an infinite potential well with $V = 0$ in the area enclosed by $y = \pm b$, $x = \pm b \pm a \cos (\pi y / b)$ and $V = \infty$ otherwise\cite{Li2002PRE,Xiong2011lpl}.
In our numerical computation,  the initial state is a moving Gaussian
wave packet and the simulation is carried out on a $64 \times 64$ grid.
The results for the entropy $S_w(\psi(t))$  are plotted in Fig. \ref{ripple} for ripple billiards
with three different values of $\epsilon = a / b$.  When $\epsilon$ is small, the system is nearly integrable and $S_w$ is almost periodic but with a decaying oscillating amplitude (see Fig.\;\ref{ripple}(a)).
As $\epsilon$ becomes larger and the system gets far away from the integrable regime,  the entropy
$S_w$ rises quickly to a maximum value and stays there with small fluctuations as discussed. The ensemble entropy  $S_E$ is also plotted and it deviates visibly
from the long-time averaged value of $S_w$. The reason is that since this is a single-particle system, $S_w$ and $S_E$ are not large. As a result, the right-hand side of the inequality (\ref{htheorem})
is not very small.

A few remarks are warranted before we conclude this section. There seems to be
a hidden assumption in von Neumann's proof of his quantum H-theorm besides
two explicitly-stated conditions (identical to conditions 1 and 2 here).
This assumption is equivalent to eigenstate thermalization
hypothesis~\cite{Srednicki1994PRE,Rigol2008Nature} as
pointed out in Ref.\cite{Srednicki2012prl} and by an anonymous referee.
In our opinion, this assumption is linked directly to  Eq.(27) in von Neumann's proof~\cite{vonNeumann2010qhtheorem}, which is highly questionable.
In contrast,  we do not have  any other assumption in our proof of the inequality
Eq. (\ref{htheorem}) besides the three conditions. The conditions for $C$ to be small,
such as the hierarchy of energy scales, have also been explicitly expressed.
Our effort here is to follow the line of von Neumann and Reimann to understand
the microscopic origin of the second law of thermodynamics without any hypothesis.
It is true that our inequality with $C\approx 0$ does not exclude the happening
of large deviation from the maximized entropy. However, this kind of large deviation
occurs rarely according to our analysis. More efforts are needed to find out exactly
how rare these events are. The usual fluctuation theorem seems not applicable here as
it depends on many concepts, such as temperature, heat bath, and entropy, whose
quantum origins are not clear themselves.

\section{Generalization to mixed states and comparison with von Neumann's entropy}
In quantum mechanics, we are all familiar with the von Neumann entropy
that is defined  as
\be
S_v(\rho) \equiv -\mathrm{tr}\; \rho \ln \rho\,,
\ee where $\rho$ is the density matrix for mixed states. This entropy $S_v$
is zero for any pure state. This fact leads to a well-known dilemma:  a large system
in a pure state has zero entropy while any of its subsystems that interacts or entangles with the rest
of the system has non-zero entropy $S_v$.

To compare our entropies to $S_v$,
we need to generalize our entropy for mixed states.
There is a straightforward way to accomplish the goal: for $N$-particle mixed states $\rho^N$,
we define
\be
S_w(\rho^N) \equiv \sum_{j_1 j_2 \dots j_N} s_w(\tr \rho^N \bm{W}_{j_1} \otimes \bm{W}_{j_2} \otimes \cdots \otimes \bm{W}_{j_N})
\ee
where $s_w(\wp) = - \wp \ln \wp$.

Several basic properties that $S_w(\rho^N)$ shares with $S_v$ \cite{Wehrl1978} are listed below :
\begin{description}
  \item[Invariance] $S_w(\rho^N)$ depends on $\{w_j\}$ but not on the choice of basis in the Hilbert space. This is a result of the invariance of the trace.
  \item[Positivity] $S_w \geq 0$ since $s_w(x) \geq 0$ for $0 \leq x \leq 1$.
  \item[Concavity] For $\lambda_1, \lambda_2 > 0$, $\lambda_1 + \lambda_2 = 1$,
  \be
  S_w(\lambda_1 \rho_1^N + \lambda_2 \rho_2^N) \geq \lambda_1 S_w(\rho_1^N) + \lambda_2 S_w(\rho_2^N). \label{eq:concavity}
  \ee
  This originates from the concavity of $s_w(x)$.

  \item[Additivity]
  \be
  S_w(\rho_1^M \otimes \rho_2^N) = S_w(\rho_1^M) + S_w(\rho_2^N) \label{eq:additivity}
  \ee
  The equality indicates that for two independent systems, the total entropy is the sum of the two. The property is also inherited from $s_w(x)$.
\end{description}
Proof of these properties is essentially the same as that in \cite{Wehrl1978} and hence omitted here.

Despite these similarities, there is one crucial difference between our entropy and $S_v$.
As we have mentioned, for $S_v$ there is a well-known dilemma: for a large system on a pure state, $S_v = 0$
while its  subsystem has non-zero entropy. In stark contrast, as we shall show, for our entropy $S_w$,
a large system always has bigger entropy than its subsystem. To demonstrate this, we only need to prove
that the entropy $S_w$ decreases when one particle is traced out of an $N$-particle system.

Without loss of generality, we tend to trace out the $N^{\textrm{th}}$ particle and write
\be
\rho^N = \sum_{i, i_N, i', i'_N} c_{i, i_N, i', i'_N} |\psi_i^{N-1},\psi_{i_N}\rangle
\langle\psi_{i'}^{N-1},\psi_{i'_N}|\,.
\ee
Here $\psi$'s are general orthonormal basis, not energy eigenstates.
With the use of the inequality $s_w(\wp_1 + \wp_2) \leq s_w(\wp_1) + s_w(\wp_2)$ for $\wp_1, \wp_2 \geq 0$,
the proof is straightforward.
\ba
&&S_w(\rho^N) \nonumber\\
&&= \sum_{j, j_N} s_w(\sum c \langle\psi_{i'}^{N-1}|\bm{W}_j^{N-1}|\psi_i^{N-1}\rangle\langle\psi_{i'_N}|\bm{W}_{j_N}|\psi_{i_N}\rangle)\nonumber\\
&&\geq \sum_{j} s_w(\sum_{j_N} \sum c \langle\psi_{i'}^{N-1}|\bm{W}_j^{N-1}|\psi_i^{N-1}\rangle\langle\psi_{i'_N}|\bm{W}_{j_N}|\psi_{i_N}\rangle)\nonumber\\
&&= \sum_{j} s_w( \sum c \langle\psi_{i'}^{N-1}|\bm{W}_j^{N-1}|\psi_i^{N-1}\rangle\langle\psi_{i'_N}|\sum_{j_N}\bm{W}_{j_N}|\psi_{i_N}\rangle)\nonumber\\
&&=\sum_{j} s_w( \sum c \langle\psi_{i'}^{N-1}|\bm{W}_j^{N-1}|\psi_i^{N-1}\rangle\langle\psi_{i'_N}|\psi_{i_N}\rangle)\nonumber\\
&&= S_w(\rho^{N-1})
\ea
where the subscripts $i, i_N, i', i'_N$ are omitted for brevity without causing confusion.

We introduce a density matrix
\be
\rho_{\rm mc}=\sum_\alpha |\braket{\psi(0)|\phi_\alpha}|^2\ket{\phi_\alpha}\bra{\phi_\alpha}\,.
\label{mc}
\ee
This density matrix can be regarded as a micro-canonical ensemble for two reasons:
(1) It is easy to check that $S_E=S_w(\rho_{\rm mc})$.
This means that the system's entropy is essentially given by $\rho_{\rm mc}$ at equilibrium.
(2) Reimann~\cite{reimann2008experimental} has also shown that the expectation of
all observables can be also be computed with $\rho_{\rm mc}$ at equilibrium.
This ensemble  is clearly different from the conventional micro-canonical ensemble in
textbooks~\cite{HuangBook} as it depends on the initial condition.
The ensemble in Eq.\;(\ref{mc}) is also different from von Neumann's\cite{Von1929} that involves certain coarse-graining of energy. However,
both our ensemble and  von Neumann's depend more or less on the choice of initial conditions.
This can lead to very interesting new physics: we are at liberty to choose  an initial condition that
composes of energy-eigenstates from two very different energy shells, which can   lead to
an equilibrium state with two distinct temperatures\cite{zhuang2014}.

\section{Conclusion}
 In summary, we have used Kohn's method to construct  a complete set of Wannier
 functions which are localized at both given positions and momenta.
 We then established a quantum phase space, where each Planck cell is represented by
one of  these Wannier functions. By mapping unitarily a quantum pure state to
this quantum phase space, we have defined an entropy for pure states.
A hierarchy of energy scales is proposed and the properties of this entropy have been examined. In particular, we have shown that
for our entropy, a system always has larger entropy than its subsystems.

The long-time dynamical behavior of our entropy has been examined and found to obey
an inequality, which like the quantum H-theorem proved by von
Neumann\cite{Von1929}, along with reasonable hypotheses, indicates that majority of
isolated quantum systems equilibrate dynamically: starting with reasonable initial states,
the quantum system will evolve into a state with maximized entropy and stay there almost
all the time with small fluctuations. Due to the time reversal symmetry, the system does
sometimes undertake  large fluctuations. However, the quantum H-theorem demands that
these large fluctuations happen rarely and are short-lived, which provides a quantum perspective of the second law of thermodynamics.

As already pointed out in the introduction, there have been renewed interests in
the foundation of quantum statistical mechanics. These new efforts have not only led to
better theoretical understanding of the issue but also to new physical predications and
challenges that await for answers from experimentalists.
For example, a quantum state which is at equilibrium but with multiple temperatures
was predicted based on the micro-cannonical ensemble established by von Neumann\cite{zhuang2014}.
And it was shown recently\cite{Goldstein2014short,Monnai2014,Short2014Rapid} that quantum
systems can relax much faster than what has been observed in reality.
Can this multiple temperature state be realized in experiments? Does it really exist
a quantum state that can relax as fast as what the theorists have predicted? The answers may
ultimately lie in understanding the borderline between the microscopic and the macroscopic
world.

\section{Acknowledgments}
We thank Hongwei Xiong and Michael Kastner for helpful discussion.
This work is supported by the NBRP of China (2013CB921903,2012CB921300) and
the NSF of China (11274024,11334001).

\appendix
\section{Proposition for orthogonalization}
We prove here a proposition based on which one can show the orthogonality of $\{w_j(x)\}$.
\newtheorem{thm1}{Proposition}
    \begin{thm1}\label{thm1}
        Assume $f_j(k) \in L^2(\mathbb{R})$ ($j \in \mathbb{N}$) if for almost every $k \in [0, 2 \pi)$,
        for all $j, j' \in \mathbb{N}$
        \be
        \sum_{n \in \mathbb{Z}} f_j^*(k + 2 n  \pi) f_{j'}(k + 2 n \pi) = \frac{1}{2 \pi} \delta_{jj'}\,,
        \ee
        then we have
        \be
        \int_\mathbb{R} f_j^*(k) f_{j'}(k) \mathrm{e}^{-i j_x k} \;\mathrm{d} k = \delta_{jj'} \delta_{0j_x}
        \ee
        for all $j, j' \in \mathbb{N}$ and $j_x \in \mathbb{Z}$. 
    \end{thm1}
    \begin{proof}
        \begin{align*}
            \int_\mathbb{R} &f_j^*(k) f_{j'}(k) \mathrm{e}^{-i j_x k} \;\mathrm{d} k\\ &= \sum_{n \in \mathbb{Z}} \int_{2 n \pi}^{2 (n + 1) \pi} f_j^*(k) f_{j'}(k) \mathrm{e}^{-i j_x k} \;\mathrm{d} k\\
            & = \sum_{n \in \mathbb{Z}} \int_0^{2 \pi} f_j^*(k + 2 n \pi) f_{j'}(k + 2 n \pi) \mathrm{e}^{-i j_x (k + 2 n \pi)} \;\mathrm{d} k\\
            & = \int_0^{2 \pi} \sum_{n \in \mathbb{Z}} f_j^*(k + 2 n \pi) f_{j'}(k + 2 n \pi) \mathrm{e}^{-i j_x k} \;\mathrm{d} k\\
            & = \frac{\delta_{jj'}}{2\pi} \int_0^{2 \pi} \mathrm{e}^{-i j_x k} \;\mathrm{d} k = \delta_{jj'} \delta_{0j_x}
        \end{align*}
    \end{proof}

The converse is also valid and the proof is omitted here.
For  the construction of localized orthonormal basis in the main text, $f_j(k) = \tilde{\w}_{j_k}(k)$.  Noticing that
\be
\int_\mathbb{R} \mathcal{F}\{f(x)\}^*(k) \mathcal{F}\{g(x)\}(k) \;\mathrm{d} k =
\int_\mathbb{R} f^*(x) g(x) \;\mathrm{d} x
\ee
and $\mathcal{F}\{f(x - j_x)\} = \mathcal{F}\{f(x)\} \mathrm{e}^{- i j_x k}$, where
\be
\mathcal{F}\{f(x)\}(k) \equiv \frac{1}{\sqrt{2 \pi}} \int_\mathbb{R} f(x) \mathrm{e}^{-i k x} \;\mathrm{d} x\,,
\ee
we have
 \ba
 &&\int_\mathbb{R} \w_{j_k}^*(x) \w_{j'_k}(x - j_x) \;\mathrm{d} x \nonumber\\
 &=& \int_\mathbb{R} \mathcal{F}\{\w_{j_k}(x)\}^*(k) \mathcal{F}\{\w_{j'_k}(x - j_x)\}(k) \;\mathrm{d} k\nonumber\\
 &=&\int_\mathbb{R} \tilde{\w}_{j_k}^*(k) \tilde{\w}_{j'_k}(k) \mathrm{e}^{-i j_x k}\;\mathrm{d} k=
 \delta_{j_k j'_k} \delta_{0j_x}\,.
 \ea

\section{Strong uncertainty relation}
The numerical results in main text with Gaussian wave packets as the initial non-orthogonal basis
indicate that $\Delta x_j\cdot\Delta k_j$ diverge at large $k$.
In fact  this divergence does not depend on the choice of initial wave packets; it
is always the case as long as one uses Kohn's method to generate a complete set of basis
that has translational symmetry.  Here we offer the proof.

Let $\{\w_{j_k}(x - j_x)\}_{j_k, j_x \in \mathbb{Z}}$ be an orthonormal basis generated from initial wave packets $\tilde{g}_{j_k}$ as discussed in section II. For convenience, we apply Schidmit orthogonalization process in order $j_k = 0, 1, -1, 2, -2, \dots$ First we show that $h(k) \equiv \lim_{j_k \rightarrow +\infty} \tilde{w}_{j_k}(k + 2 j_k \pi)$ exists. Since $\tilde{\w}_{j_k}(k + 2 n \pi) \equiv v_{k, j_k}(n) / \sqrt{2 \pi}$, it is sufficient to prove that for any fixed $k$, $v_k(n) \equiv \lim_{j_k \to +\infty} v_{k, j_k}(n + j_k)$ exists. Introduce a subspace of $l^2(\mathbb{Z})$
\be
H_{k, l} \equiv \spp_{|j_k| \leq l} v_{k, j_k}(n + l + 1)
\ee
for $l \in \mathbb{N}$. Noticing $v_{k, j_k}$ is the linear combination of $u_{k, j'_k}$ ($|j'_k| \leq |j_k|$), \ba
&H_{k, l} &= \spp_{|j_k| \leq l} u_{k, j_k}(n + l + 1) \nonumber \\
&&= \spp_{|j_k| \leq l} \tilde{g}_{j_k}(k + 2 (n + l + 1) \pi) \nonumber \\
&&= \spp_{|j_k| \leq l} \tilde{g}_0(k + 2 (n + l - j_k + 1) \pi) \nonumber \\
&&= \spp_{1 \leq j_k \leq 2 l + 1} \tilde{g}_0(k + 2 j_k \pi + 2 n \pi)
\ea
From Schidmit orthogonalization process, we know that
\ba
&v_{k, l+1}(n + l + 1) &= \bm{N} \bm{P}_{H_{k, l}^\bot} u_{k, l+1}(n + l + 1)\nonumber\\
&& =  \bm{N} \bm{P}_{H_{k, l}^\bot} \tilde{g}_0(k + 2 n \pi)
\label{eq:B3}
\ea
where $\bm{N}$ is the normalization operator and $\bm{P}$ is the projection operator. We have to prove that as $l \to +\infty$, the limit of (\ref{eq:B3}) exists. Since $\bm{N}$ is continuous except at the origin, it is sufficient to show that
\ba
\lim_{l \to +\infty} \bm{P}_{H_{k, l}^\bot} \tilde{g}_0(k + 2 n \pi) \;\textrm{exists and} \; \neq 0.
\ea

Denote $r_{k, l} \equiv \bm{P}_{H_{k, l}^\bot} \tilde{g}_0(k + 2 n \pi)$. Since $H_{k, l}^\bot$ is monotonically decreasing and $\bm{P}$ is orthogonal projection, $\|r_l\|$(subscript $k$ omitted) is decreasing and $r_l - r_{l'} \bot r_{l'}$ for $l < l'$. Hence $\|r_l - r_{l'}\|^2 = \|r_l\|^2 - \|r_{l'}\|^2 \to 0$ as $l \to +\infty$. $\{r_l\}$ converges.

Note that
\be
r_k \equiv \lim_{l \to +\infty} r_{k, l} \neq 0 \Leftrightarrow \tilde{g}_0(k + 2 n \pi) \notin H_k \equiv \overline{\bigcup_{l \in \mathbb{N}}
H_{k, l}}.
\ee
For reasonable initial wave packets, this is always the case. For example, when $\tilde{g}_0$ is compactly supported, $\exists N$ such that $\tilde{g}_0(k + 2 N \pi) \neq 0$ and $\forall n > N$, $\tilde{g}_0(k + 2 n \pi) = 0$, hence $\tilde{g}_0 \notin H_{k}$.

From the proof above, we see $v_{k, l+1} \in H_{k, l} ^\bot$, hence, $v_{k, l + 1} \bot v_{k, j_k}$ for all $|j_k| \leq l$. As a result of the limiting process, $r_k(n) \bot r_k(n + p)$ for all $p \neq 0$. The proof is valid for every $k$, hence $h(k)$ and its translations form a periodic orthonormal basis. Due to Balian's proof\cite{bourgain1988remark}, for $h$, $\Delta x = \infty$ or $\Delta k = \infty$. Thus, no matter how the initial wave packet is chosen, by our orthogonalization approach, strong uncertainty relation holds, if space translational symmetry is required, i.e. $$\sup\Delta x_j \sup\Delta k_j = \infty.$$

\section{Proof of the inequality}
Here we present the detailed proof of the inequality (\ref{htheorem}).
First we provide two inequalities that will be useful later:
\be
\overline{\left(\wp_j(t) - \overline{\wp}_j\right)^2}\leq \overline{\wp}_j^2\,,
\label{ineq:2}
\ee
and
\begin{align}
    \overline{\left(\wp_j(t) - \overline{\wp}_j\right)^4} \leq 9 \overline{\wp}_j^4\,.
    \label{ineq:4}
\end{align}
Derivation of inequality (\ref{ineq:2}) needs conditions 1 and 2 while for inequality (\ref{ineq:4}) all three conditions are necessary.
The proof of the inequality (\ref{ineq:2}) is as follows.
\begin{align}
    \overline{\left(\wp_j(t) - \overline{\wp_j}\right)^2} &=
    \sum_{\alpha \neq \beta} c_{\alpha j} c_{\alpha j}^* c_{\beta j} c_{\beta j}^* \nonumber \\&=
    \overline{\wp}_j^2 - \sum_\alpha |c_{\alpha j}|^4 \leq \overline{\wp}_j^2\,.
    \label{ineq:sqr}
\end{align}

It is straightforward but more care is needed  to prove (\ref{ineq:4}). Since in the proof, only one Planck cell $j$ is considered, we suppress the subscript $j$ of $c_{\alpha j}$ for brevity.
Before time averaging, we have
\ba
&&\left(\wp_j(t) - \overline{\wp}_j\right)^4
=\big(\sum_{\alpha \neq \alpha'}c_{\alpha} c_{\alpha'}^*\mathrm{e}^{i(E_\alpha-E_{\alpha'})t}\big)^4\nonumber\\
&=& \sum_{\alpha \neq \alpha', \beta \neq \beta'}^{\chi \neq \chi', \gamma \neq \gamma'}
c_{\alpha} c_{\alpha'}^* c_{\beta} c_{\beta'}^*c_{\chi}^* c_{\chi'}
c_{\gamma}^* c_{\gamma'} \nonumber\\
&&\times~\mathrm{e}^{i(E_{\alpha} + E_{\beta} + E_{\chi'} + E_{\gamma'} -
E_{\alpha'} - E_{\beta'} - E_{\chi} - E_{\gamma})t} \,.
\label{eq:4}
\ea
This yields
\ba
&&\overline{\left(\wp_j(t) - \overline{\wp}_j\right)^4}=\nonumber\\
&& \overline{\sum}_{\alpha \neq \alpha', \beta \neq \beta'}^{\chi \neq \chi', \gamma \neq \gamma'}
c_{\alpha} c_{\alpha'}^* c_{\beta} c_{\beta'}^*c_{\chi}^* c_{\chi'}
c_{\gamma}^* c_{\gamma'} \,,
\ea
where the overlined summation is  over only the terms that satisfy the energy relation
\be
E_{\alpha'} + E_{\beta'} + E_{\chi} + E_{\gamma}=
E_\alpha + E_\beta + E_{\chi'} + E_{\gamma'}\,.
\ee
This sum can be divided into four parts.
\begin{itemize}
  \item $\alpha = \beta'$ and $\beta = \alpha'$\,.\\
In this case, the energy relation becomes $ E_{\chi} + E_{\gamma}=
 E_{\chi'} + E_{\gamma'}$. According to condition 2, $\chi=\gamma'$
 and $\gamma=\chi'$. The sum of relevant terms converts into
    \ba
     &&\sum_{\alpha \neq\beta, \chi \neq \gamma}
 c_{\alpha} c_{\beta}^* c_{\beta} c_{\alpha}^*c_{\chi}^* c_{\gamma}
 c_{\gamma}^* c_{\chi}
 \leq \overline{\wp}_j^4\,.
    \ea
\item $\alpha = \beta'$ and $\beta \neq \alpha'$ \,.\\
The energy relation is now
\be
E_{\chi} + E_{\gamma}-E_\beta=E_{\chi'} + E_{\gamma'}-E_{\alpha'}\,.
\ee
({\it i}) When $\chi=\beta$, $E_{\gamma}- E_{\gamma'}=E_{\chi'} -E_{\alpha'}$.
With condition 2, this implies that $\gamma=\chi'$ and $\gamma'=\alpha'$.
We then have
\ba
    &&\sum_{\alpha \neq \alpha', \beta \neq \alpha}^{\beta\neq\gamma,\gamma \neq \alpha'}
c_{\alpha} c_{\alpha'}^* c_{\beta} c_{\alpha}^*c_{\beta}^* c_{\gamma}
c_{\gamma}^* c_{\alpha'} \leq \overline{\wp}_j^4\,.
\ea
({\it ii}) When $\chi\neq\beta$, we have $\gamma=\beta$. This can be seen in the energy relation $E_{\chi} + E_{\gamma}-E_\beta-E_\delta=E_{\chi'} + E_{\gamma'}-E_{\alpha'}-E_\delta$ with $E_\delta$ being
an arbitrary eigen-energy. If $\gamma \neq \beta$, choose $\delta$ such that $\{\chi, \gamma\}\cap\{\beta, \delta\} = \emptyset$; according to condition 3, it is required $\beta = \alpha'$ which is a contradiction. Since $\gamma = \beta$, the rest calculation is similar to that in ({\it i}).

Overall, this part of summation is no more than $2\overline {\wp}_j^4$.

 \item $\alpha \neq \beta'$ and $\beta = \alpha'$\,.\\
 Similarly, this part contributes $2\overline {\wp}_j^4$.

  \item $\{\alpha, \beta\} \cap \{\alpha', \beta'\} = \emptyset$\,.\\
In this case, condition 3 demands that $\{\alpha,\beta\}=\{\chi,\gamma\}$ and
$\{\alpha',\beta'\}=\{\chi',\gamma'\}$. Among the  four different combinations,
we choose $\alpha=\chi, \beta=\gamma, \alpha'=\gamma', \beta'=\chi'$, which leads to
\ba
&&\sum_{\alpha \neq \alpha', \gamma \neq \beta'}^{\gamma \neq \alpha'}
c_{\alpha} c_{\alpha'}^* c_{\gamma} c_{\beta'}^*c_{\alpha}^* c_{\beta'}
c_{\gamma}^* c_{\alpha'}\leq \overline {\wp}_j^4 \,.
\ea
As the three other combinations are similar, the sum is less or equal to $4\overline {\wp}_j^4$.

\end{itemize}
Summing all these cases, we obtain the inequality (\ref{ineq:4}).

In the proof, we also use the following equality: for $\wp, \overline{\wp} \geq 0$, there exists $0 < \xi \leq 1$ such that
\be
\wp \ln \wp = \overline{\wp} \ln \overline{\wp} + (1 + \ln \overline{\wp}) (\wp - \overline{\wp}) + \frac{\xi(\wp - \overline{\wp})^2}{\overline{\wp}}\,.
\ee

We are now ready to present the full  proof of the inequality (\ref{htheorem}). Assume for every $j$, $\overline{\wp}_j < 1 / \mathrm{e}$.

\begin{widetext}
\begin{align*}
    &\overline{\left(\sum_j \left[\wp_j(t) \ln \wp_j(t) - \overline{\wp}_j \ln \overline{\wp}_j\right]\right)^2}
    = \overline{\left(\sum_j\left[(1 + \ln \overline{\wp}_j) (\wp_j(t) - \overline{\wp}_j) + \frac{ \xi_j(t)(\wp_j(t) - \overline{\wp}_j)^2}{\overline{\wp}_j}\right]\right)^2}\\
    = & \sum_{j, j'} (1 + \ln \overline{\wp}_j) (1 + \ln \overline{\wp}_{j'}) \overline{ (\wp_j(t) - \overline{\wp}_j) (\wp_{j'}(t) - \overline{\wp}_{j'})}
    + 2\sum_{j, j'} \frac{1 + \ln \overline{\wp}_j}{\overline{\wp}_{j'}} \overline{(\wp_j(t) - \overline{\wp}_j) \xi_{j'}(t)(\wp_{j'}(t) - \overline{\wp}_{j'})^2}\\
    &\quad+ \sum_{j, j'} \frac{1}{\overline{\wp}_j \;\overline{\wp}_{j'}} \overline{ \xi_{j}(t) \xi_{j'}(t)(\wp_j(t) - \overline{\wp}_j)^2 (\wp_{j'}(t) - \overline{\wp}_{j'})^2} \\
    \leq &\sum_{j, j'} (1 + \ln \overline{\wp}_j) (1 + \ln \overline{\wp}_{j'}) \overline{\wp}_j \;\overline{\wp}_{j'} C_{j j'}
    - 2\sum_{j, j'} \frac{1 + \ln \overline{\wp}_j}{\overline{\wp}_{j'}} \sqrt{\overline{(\wp_j(t) - \overline{\wp}_j)^2} \; \overline{ (\wp_{j'}(t) - \overline{\wp}_{j'})^4}}\\
    &\quad+ \sum_{j, j'} \frac{1}{\overline{\wp}_j \;\overline{\wp}_{j'}} \sqrt{\overline{(\wp_j(t) - \overline{\wp}_j)^4}\;\overline{(\wp_{j'}(t) - \overline{\wp}_{j'})^4}}\\
    \leq & \sum_{j, j'} (1 + \ln \overline{\wp}_j) (1 + \ln \overline{\wp}_{j'}) \overline{\wp}_j \;\overline{\wp}_{j'} C_{j j'} - 2 \sum_{j, j'}\frac{1 + \ln \overline{\wp}_j}{\overline{\wp}_{j'}} \overline{\wp}_j \;3 \overline{\wp}_{j'}^2
    + \sum_{j, j'} \frac{9}{\overline{\wp}_j \;\overline{\wp}_{j'}} \overline{\wp}_{j}^2 \overline{\wp}_{j'}^2 \\
    \leq& C S_E^2   + 8 S_E + 4\,.
    \end{align*}
\end{widetext}
For the first inequality ``$\le$", we have used $\overline{\wp}_j < 1 / \mathrm{e}$; in the last step, we have used $|C_{j j'}| \leq 1$.  For a typical wave function of a many-body quantum system,
it spreads out over thousands of Planck cells in the phase space; therefore
$\overline{\wp}_j < 1 / \mathrm{e}$ is satisfied almost always.

\section{Estimate of $S_E$ and $C$}


\paragraph*{Estimate of $S_E$} This is equivalent to
show Eq.\;(\ref{SEmax}).

Suppose that $J \equiv [E, E + \Delta E]$ ($\RE \ll \Delta E \lesssim \deltaE$) is a macroscopic energy shell which is significantly occupied by
the quantum state $\psi$. For brevity, we use $j\in J$ to represent
$E_j=\braket{w_j|H|w_j}\in [E, E + \Delta E]$,
and $\alpha\in J$ for $E_\alpha\in [E, E + \Delta E]$.

As $\RE$ is the correlation energy scale,  we have
\be
\wp_\alpha |\braket{w_j|\phi_\alpha}|^2\ll 1\,,
\ee
when $|E_j-E_\alpha|\gg\RE$. Since $\Delta E\gg\RE$, we have for a typical (energy away from end points of $J$) $j \in J$
\be
\overline{\wp}_j \approx \sum_{\alpha \in J} \wp_\alpha |\la w_j | \phi_\alpha \ra|^2\,,
\ee
and for a typical $\alpha \in J$
\be
\sum_{j \in J} |\la w_j | \phi_\alpha \ra|^2 \approx 1\,.
\ee
With (\ref{hierarchy}) and (\ref{inevitable}), these lead to
\be
\sum_{j \in J} \overline{\wp}_j \approx \sum_{j \in J} \sum_{\alpha \in J} \wp_\alpha |\la w_j | \phi_\alpha \ra|^2 \approx \sum_{\alpha \in J} \wp_\alpha \,.
\ee

For matrix $A_{j \alpha} \equiv |\la w_j | \phi_\alpha \ra|^2 $($j \in J$, $\alpha \in J$),
the sum of every row of $A^T A$ is less than one. By the Perron-Frobenius theorem, the eigenvalue of $A^T A$ must be equal or less than one in module and we have
\be
\sum_{j \in J} \overline{\wp}_j^2 \approx P^T A^T A P \leq P^T P
= \sum_{\alpha \in J} \wp_\alpha^2\,,
\ee
where the column vector $(P)_\alpha = \wp_\alpha$.  We are now ready to estimate
the entropy,
\ba
S_{E,J} &\equiv& -\sum_{j \in J} \overline{\wp}_j \ln \overline{\wp}_j \geq -\sum_{j \in J} \overline{\wp}_j \ln\big(\sum_{j \in J} \overline{\wp}_j^2 \big/ \sum_{j \in J} \overline{\wp}_j \big) \nonumber \\
&\gtrsim& -\sum_{\alpha \in J} \wp_\alpha \ln\big(\sum_{\alpha \in J} \wp_\alpha^2  \big/  \sum_{\alpha \in J} \wp_\alpha \big) \nonumber\\
&\approx& -\sum_{\alpha \in J} \wp_\alpha \ln\big( \la \wp_\alpha\ra_s  / R \big) \nonumber\\
&\approx& S_{E,J}^{\max} + \sum_{\alpha \in J} \wp_\alpha \ln R \,. \quad\label{D7}
\ea
In the first line of the above derivation, Jensen's inequality for function $- \ln x$ is applied.
If $J$ is the only energy shell significantly occupied by $\psi$, we already have Eq. (\ref{SEmax}).
If we have more than one such energy shells, we
sum Eq.\;(\ref{D7}) for all $J$ and obtain Eq. (\ref{SEmax}).\\



\paragraph* {Estimate of $C$} Define $d_{\epsilon, j}$ as the minimal $d$ such that
\be\sum_{|E_\alpha - E_j| < d} \wp_\alpha |\la w_j | \phi_\alpha \ra|^2 > (1 - \epsilon) \overline{\wp}_j \label{defined}\,,
\ee
where $E_j$ is the average energy of $w_j$.  We say that Planck cells $j$ and $j'$ overlap
if $|E_j - E_{j'}| < d_{\epsilon, j} + d_{\epsilon, j'}$. We consider two cells $j$ and $j'$
which do not overlap.  With condition 1 and 2, we have
\be
C_{j j'} \overline{\wp}_j \overline{\wp}_{j'} \leq \big|\sum_\alpha \wp_\alpha \la w_j | \phi_\alpha \ra \la \phi_\alpha | w_{j'} \ra \big|^2\,.
\ee
Without loss of generality, we assume $E_j < E_{j'}$ and split the above sum
into two parts: one part close to $j$ and the other close to $j'$.
\ba
C_{j j'} \overline{\wp}_j \overline{\wp}_{j'} \leq 2\big|\sum_{E_\alpha < E'} \wp_\alpha \la w_j | \phi_\alpha \ra \la \phi_\alpha | w_{j'} \ra \big|^2 \nonumber\\+  2\big|\sum_{E_\alpha \geq E' } \wp_\alpha \la w_j | \phi_\alpha \ra \la \phi_\alpha | w_{j'} \ra \big|^2  \,,
\ea
where $E' \equiv E_j + d_{\epsilon, j}$. The first term, by Cauchy-Schwartz inequality, is less or equal to
\be
2 \sum_{E_\alpha < E'} \wp_\alpha |\la w_j | \phi_\alpha \ra|^2\sum_{E_\alpha < E'} \wp_\alpha |\la \phi_\alpha | w_{j'} \ra |^2 < 2 \epsilon \;\overline{\wp}_j \overline{\wp}_{j'}\,.
\ee
Similar argument applies to the second term and we have
\be
C_{j j'} \leq 4 \epsilon\,.
\ee




Next we need to establish that non-overlapping Planck cells are the majority in the pair of cells involved. For this purpose,
it is sufficient to show that for any significantly occupied Planck cell $j$, the sum $S_j$ of entropies over cells overlapping with $j$ is much less than $S_E$. This is indeed the case,
\ba
S_j&\lesssim&-\int_{E_j - d_\epsilon}^{E_j + d_\epsilon} \d E\; \rho(E) \la \wp_\alpha \ra_s(E) \ln \la \wp_\alpha \ra_s(E) \qquad\nonumber \\
&\ll& -\int_{E_j - \deltaE}^{E_j + \deltaE} \d E\; \rho(E) \la \wp_\alpha \ra_s(E) \ln \la \wp_\alpha \ra_s(E) \nonumber \\
&\leq& S_E^{\max} \approx S_E\,.
\ea
In the above derivation, we have used that $\la \wp_\alpha \ra$ are effectively constant on a range of $\deltaE$, $S_E$ is maximized, and assumption (\ref{inevitable}). This yields the inequality (\ref{estimateC}) if we set $\epsilon=\RE/\deltaE$ and hence $\RE \ll d_\epsilon \lesssim \sqrt{\RE \deltaE} \ll \deltaE$ (estimated according to Eq.\;(\ref{RE}) and (\ref{defined}) for typical $j$).


\end{document}